\documentclass[a4paper]{jpconf}
\usepackage{graphicx}
\begin{document}
\title{Oscillation modes of strange quark stars with a strangelet crust}

\author{Jessica Asbell and Prashanth Jaikumar}

\address{Department of Physics \& Astronomy, California State University Long Beach, 1250 Bellflower Blvd., Long Beach, CA 90840 USA}

\ead{prashanth.jaikumar@csulb.edu}

\begin{abstract}
We study the non-radial oscillation modes of strange quark stars with a homogeneous core and a crust made of strangelets. Using a 2-component equation-of-state model (core+crust) for strange quark stars that can produce stars as heavy as 2 solar masses, we identify the high-frequency l=2 spheroidal ($f$, $p$) in Newtonian gravity, using the Cowling approximation. The results are compared to the case of homogeneous compact stars such as polytropic neutron stars, as well as bare strange stars. We find that the strangelet crust only increases very slightly the frequency of the spheroidal modes, and that Newtonian gravity overestimates the mode frequencies of the strange star, as is the case for neutron stars.

\end{abstract}

\section{Introduction}
The recent direct detection of gravitational waves from the inspiral and merger of a pair of massive Black Holes by Advanced LIGO heralds a new era of compact star observations~\cite{Abbott:2016blz,Abbott:2016nmj}. It is hoped that systematic analysis of Advanced LIGO data will be able to pin down the equation of state of dense matter relevant to compact  stars. The merger and ring down phases of colliding neutron star and black-hole binaries~\cite{Creighton:2003nm,Bauswein:2012ya} are the strongest transient events, but they leave in their wake many different oscillations of the stellar fluid that carry information on the equation of state of dense matter~\cite{Kokkotas:1999bd,Kokkotas:1999mn}. The classification of these modes is a well-developed field, and their study is promising because the spectrum of such modes is connected to underlying stellar structure~\cite{PenaGaray:2008qe,Sotani:2012qc}. The even-parity or spheroidal modes result from density and pressure perturbations to the star, while the odd-parity axial modes are non-trivial only for rotating stars. The spectrum of spheroidal $f,p,g$ modes was recently studied by Lugones and Vasquez-Flores~\cite{Flores:2013yqa} for stars made of hadronic, self-bound and hybrid matter, in order to find discriminating features among them. Other works have also discussed the differences in mode frequencies between neutron stars and quark/hybrid stars~\cite{Kojima:2002iv,Moraes:2014dra,Chatziioannou:2015uea,Fu:2017mcw}. The work reported in these proceedings considers the effect of a crust made of non-superconducting quark matter on the spectrum of spheroidal modes in a strange star. Our 2-component model for strange stars includes a crust made of strangelets~\cite{Jaikumar:2005ne,Alford:2006bx} and therefore differs in an essential way from previous studies, where only homogeneous quark phases or mixed phases of quarks and hadronic matter~\cite{Sotani:2012ej} were considered.

\section{The 2-Component Model for Strange Quark Stars}
The term quark star (QS), or strange quark star (SQS), commonly refers
to a compact star composed completely of self-bound strange quark
matter, and is a possible consequence of the strange quark matter hypothesis~\cite{1984PhRvD}. One can then describe a compact star with an appropriate
equation of state (EOS) for deconfined quarks at high density, and determine if the oscillation spectra of such a star is different from that computed with a nuclear matter
EOS. 

\subsection{Core EOS} The core is assumed to consist of homogeneous and unpaired charge neutral 3-flavor interacting quark matter, which we describe using the simple thermodynamic Bag model EOS~\cite{Alford:2004pf} with ${\cal O}(m_s^4)$ corrections to account for the moderately heavy strange quark. Perturbative interactions to non-interacting quark matter~\cite{Fraga:2001id} are absorbed into a parameter $(1-a_4)\sim {\cal O}(\alpha_s^2)\approx 0.3$ as suggested in~\cite{Alford:2004pf} to extend the applicability of the model to stars as heavy as $\approx 2M_{\odot}$. The core EOS is
\begin{equation}
\label{eqn1}
P_\mathrm{core}=\frac{1}{3}(\epsilon-4 B)-\frac{m_s^2}{3\pi}\sqrt{\frac{\epsilon-B}{a_4}}+\frac{m_s^4}{12\pi^2}\left[2 - \frac{1}{a_4}+3\ln\left(\frac{8\pi}{3m_s^2}\sqrt{\frac{\epsilon-B}{a_4}}\right)\right] , 
\end{equation}
where $\epsilon$ is the energy density of homogeneous quark matter (also to ${\cal O}(m_s^4)$ in the Bag model) and $B$ the Bag constant which we fix by requiring that the first-order transition between neutral quark matter and the vacuum ($P$=0) occur at a quark chemical potential $\mu_q$=$\mu_{\rm crit}\leq$ 310 MeV~\cite{Alford:2006bx}. This ensures that the hypothesis of absolute stability for 3-flavor quark matter is not violated. Corrections due to the superconducting gap $\Delta$ can be included in the EOS as in~\cite{Rupak:2012wk}, but are omitted here to avoid proliferation of parameters. In Fig.1, we show the mass-radius relationship for homogeneous quark stars for different $\mu_{\rm crit}$ values at $m_s$=100 MeV, with $B$ determined from the criterion for absolute stability. \\
\begin{figure}[htbp]
\begin{center}
\label{mrbare}
\includegraphics[height=3in,width=3.5in]{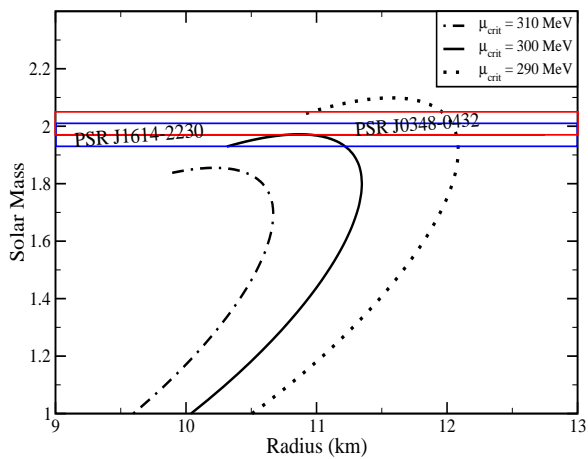}
\caption{Mass-Radius relation for homogeneous quark stars described by the EoS (Eq.~\ref{eqn1}). The interaction parameter $a_4$=0.7 and $m_s$=100 MeV. Measured limits on the mass of 
PSR J1614-2230 (1.97$\pm$0.04 $M_{\odot}$) and PSR J0348-0432 (2.01$\pm$0.04 $M_{\odot}$) are shown.}
\end{center}
\end{figure}

Usually, theoretical models of quark stars consist of entirely homogeneous stars comprised
of strange quark matter or a thin nuclear crust on top of a pure strange star. These cases
predict steep drops in density as the surface is approached and the pressure goes to zero.
They also predict large electric fields at the surface (or just below, in the case of a nuclear crust). 
The model we use differs from these in that it considers a heterogeneous crust on top of a fluid quark
star. When the condition of local charge neutrality is relaxed, quarks and electrons can form a mixed phase. This model has the advantage
that it reduces the density gradient near the crust and produces negligible electric field energy at
the starÕs surface. However, there will be short-range (screened) electric fields inside the
mixed phase, which can be tolerated if the Gibbs free energy is lowered sufficiently by
forming a mixed phase. The thickness of such a crust would be small but on the same
order as the crust of a neutron star, approximately $10^4$ cm. This amounts to no
more than 1\% of the stellar radius.

\subsection{Crust EOS} 
\begin{figure}[htb!]
\begin{center}
\includegraphics[height=1.5in,width=3.5in]{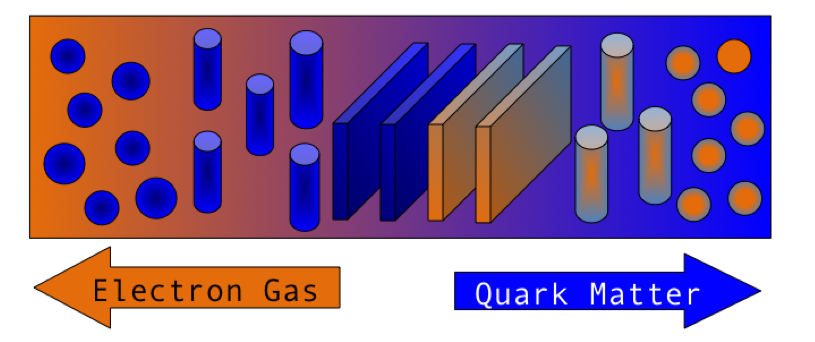}
\caption{Diagram of a mixed-phase crust with net-positive nuggets of quark matter in a neutralizing electron background
(voids). Pressure and density increase toward the center of the star, causing the nuggets to
deform and fuse, and nuggets of negatively charged electrons embed in a positive quark background (voids). Deeper into the
star, at very high pressure, quark matter becomes homogeneous.}
\label{mixed}
\end{center}
\end{figure}

We assume that the crust is in a non-superconducting, globally neutral mixed phase of strangelets and electrons, so we can choose the configuration described in~\cite{Jaikumar:2005ne} which has ungapped 3-flavor quark matter with massive strange quark. The composition of the crust changes with depth as the quark phase fraction $x$ increases from zero at the surface to one in the homogeneous phase. Phase coexistence with stable strangelets requires that (in the absence of surface tension) the pressure inside and outside the strangelet be the same, hence the quark pressure $P_q(\mu_q,\tilde{\mu}_e)$ in the mixed phase is zero, and the pressure is only due to electrons which depends on the electron chemical potential $\tilde{\mu}_e$ as
\begin{equation}
\label{pmix}
P_{\rm crust}=\frac{\tilde{\mu}_e^4}{12\pi^2}\,,\quad \tilde{\mu}_e=\frac{n_Q}{\chi_Q}\left(1-\sqrt{1-\xi}\right)\,,\quad \xi=\frac{2P_0\chi_Q}{n_Q^2}
\end{equation}
where $n_Q(\mu_q,m_s)$ and $\chi_Q(\mu_q,m_s)$ represent the quark charge and quark susceptibility (both slowly varying in the mixed phase) and $P_0$ is the pressure of quarks without electrons, which varies considerably in the mixed phase. These generic relations can be applied to a specific model of quark matter, so long as we work to second order in the (small) electron chemical potential. This is a well-justified and useful approximation since $\tilde{\mu}_e/\mu_q\approx 0.05$. Just as for the core, we adopt the Bag model EOS to describe the crust strangelets with ${\cal O}(m_s^4)$ corrections, from which it follows that to the same order: $n_Q(\mu_q)=m_s^2\mu_q/(2\pi^2)$ and $\chi_Q(\mu_q)=(2\mu_q^2)/(\pi^2$). Homogeneous quark matter gives way to the mixed phase crust at a radius $r$=$r_c$ where $\xi|_{r=r_c}$=1. The models with crust have their surface at $\mu_q=\mu_{\rm crit}$, so they have very nearly the same mass and radius as the bare strange star models. A sketch of the constituents in the mixed phase is given in Fig.\ref{mixed}.

\section{Method for Computing Oscillation Spectra} 
In order to perform a perturbative analysis of compact star oscillations, we first
must model their equilibrium structure. The
Tolman-Oppenheimer-Volkoff (TOV) equations are used to provide the equilibrium structure of a
compact star. We approach the non-radial oscillation problem in Newtonian gravity and employ the
Cowling approximation, which allows one to neglect perturbations in the gravitational
potential~\cite{Kastaun:2008jr}. Under the Cowling approximation, the metric is
stationary and matter is decoupled from space-time. While it is impossible to generate gravitational waves in Newtonian gravity, this approximation
is justified as a first approximation since it simplifies the system considerably. However, the $f$ and $p$ mode
frequencies computed from general relativistic calculations may differ significantly. To describe the fluid pulsations, we choose a reference frame in which the unperturbed fluid velocity is zero and search for oscillating solutions by writing the displacement as $\vec{u}$ = $\vec{\xi}({\bf x}){\rm e}^{i \sigma t}$ which describes a wave with frequency
$\sigma$ and amplitude $\vec{\xi}({\bf x})$. In spherical coordinates we make the radial and angular dependencies of the displacement
explicit $\vec{\xi}({\bf x})=(\xi_r,\xi_{\theta},\xi_{\phi})$ and identify the fluid pulsation variables
\begin{equation} 
     \xi_r = U(r)Y_{lm}\,, \quad
     \xi_{\theta} = V(r)\frac{\partial Y_{lm}}{\partial \theta}\,,\quad
     \xi_{\phi} = \frac{V(r)}{{\rm sin}\,\theta} \frac{\partial Y_{lm}}{\partial \phi}
\end{equation}
where $U(r)$ and $V(r)$ are the radial and transverse displacement eigenfunctions. In general, the perturbation equations we solve for a 1-component (homogeneous) star are~\cite{1988ApJ725M}:
\begin{equation}
\label{eq:14}
    (1+\widetilde{V})\frac{dy_1}{dx}=\left(\frac{\widetilde{V}}{\Gamma} - 3\right)y_1 +\left(\frac{l(l+1)}{c_1 \Omega^2} -\frac{\widetilde{V}}{\Gamma}\right)y_2
\end{equation}
\begin{equation}
\label{eq:15}
    (1+\widetilde{V})\frac{dy_2}{dx}=(c_1 \Omega^2 + \mathcal{A}r)y_1 + (1-\widetilde{U}-\mathcal{A}r)y_2
\end{equation}
with the following definitions: $y_1$=$\frac{U(r)}{r}$, $y_2$=$\frac{\delta P}{\rho g r}$, $\widetilde{V}$=$\frac{\rho g r}{p}$, $\widetilde{U}$=$\frac{d \ln(M_r)}{d \ln(r)}$, $c_1$=$\left(\frac{r}{R_{\star}}\right)^3\frac{M_{\star}}{M_r}$, $\Omega^2$=$\frac{\sigma^2 R_{\star}^3}{M_{\star}}$ where $\delta P$ is the Eulerian pressure perturbation, $M_r$ is the mass contained in radius $r$, $M_{\star}$ and $R_{\star}$ are the mass and radius of the star and $\mathcal{A}$ is the Schwarzschild gradient for local convective stability. In our analysis, we set $\mathcal{A}$=0 for the bare strange star, although finite temperature effects or composition gradients can change the situation. This system of pulsation equations is solved numerically to find the oscillation eigenfrequencies $(\sigma)$ that obey the boundary conditions that solutions $y_1$ and $y_2$ are regular at the origin ($r$=0) and the Lagrangian variation of the pressure vanishes at the surface ($r$=$R_{\star}$). In addition, the solutions are normalized such that the relative radial displacement $y_1(R_{\star})$=1. For a 2-component star, the crust eigenfunctions are connected to those in the core through suitable continuity conditions. The complete and lengthier set of 4 equations (2 for the fluid displacement and 2 for the tractions) and the relevant boundary conditions are written down in~\cite{1988ApJ725M,Yoshida:2002vd}, so we do not reproduce them here. We use these to obtain results for the 2-component strange star with a crust.

\section{Results: Oscillation Spectra for Strange Quark Stars} 

In Table~\ref{polytrope-freq}, we show the mode period values for the $l$=2 $f$, $p$ and core
$g$ modes for a neutron star modeled as a single component, fluid, non-rotating
core described by a polytropic EOS. We consider three different values of the polytropic index (n=1.0, 1.5, 2.0), keeping stellar mass and radius fixed at $M$=1.4$M_{\odot}$ and $R$=10 km. Mode frequencies and periods are related in the usual way, f=1/$\Pi$. The $f$ modes are short period, corresponding to a few kHz and the $p$ modes are higher in frequency still. The core $g$ modes  are in the range of a few Hz to few tens of Hz. This is as analytically predicted by Cox~\cite{Cox} and the trend with increasingly relativistic models is as found by Andersson \& Kokkotas~\cite{Andersson:1997rn} and McDermott et al~\cite{1988ApJ725M}, although their exact values are somewhat different due to different choices of neutron star model. For neutron stars, the $f$ mode frequency increases as the square root of the mean density as expected, and hence also stellar mass.

\begin{table}[htp]
\caption{The $f,p,g$ oscillation mode frequencies for a single-component neutron star are listed below.  $n$ refers to the polytropic index and azimuthal mode number $l=2$.  The mass is fixed at $1.4~M_{\odot}$ and radius at $10$~km in all cases.}
\vskip 0.5cm
\begin{center}
\begin{tabular}{|c|c|c|c|}
\hline
 $l$ = 2 & $f-$mode [kHz] & $p_1-$mode [kHz]& $g_1-$mode [Hz] \\
 \hline
 $n=1.0$   & 3.59    & 8.23    & 4.63  \\
 $n=1.5$   & 4.28    & 9.85    & 13.4  \\
 $n=2.0$   & 5.11    & 11.5    & 24.4  \\
 \hline
 \end{tabular}
\end{center}
\label{polytrope-freq}
\end{table}

\begin{figure}[h]
\begin{minipage}{18.8pc}
\includegraphics[width=19.8pc,height=15pc]{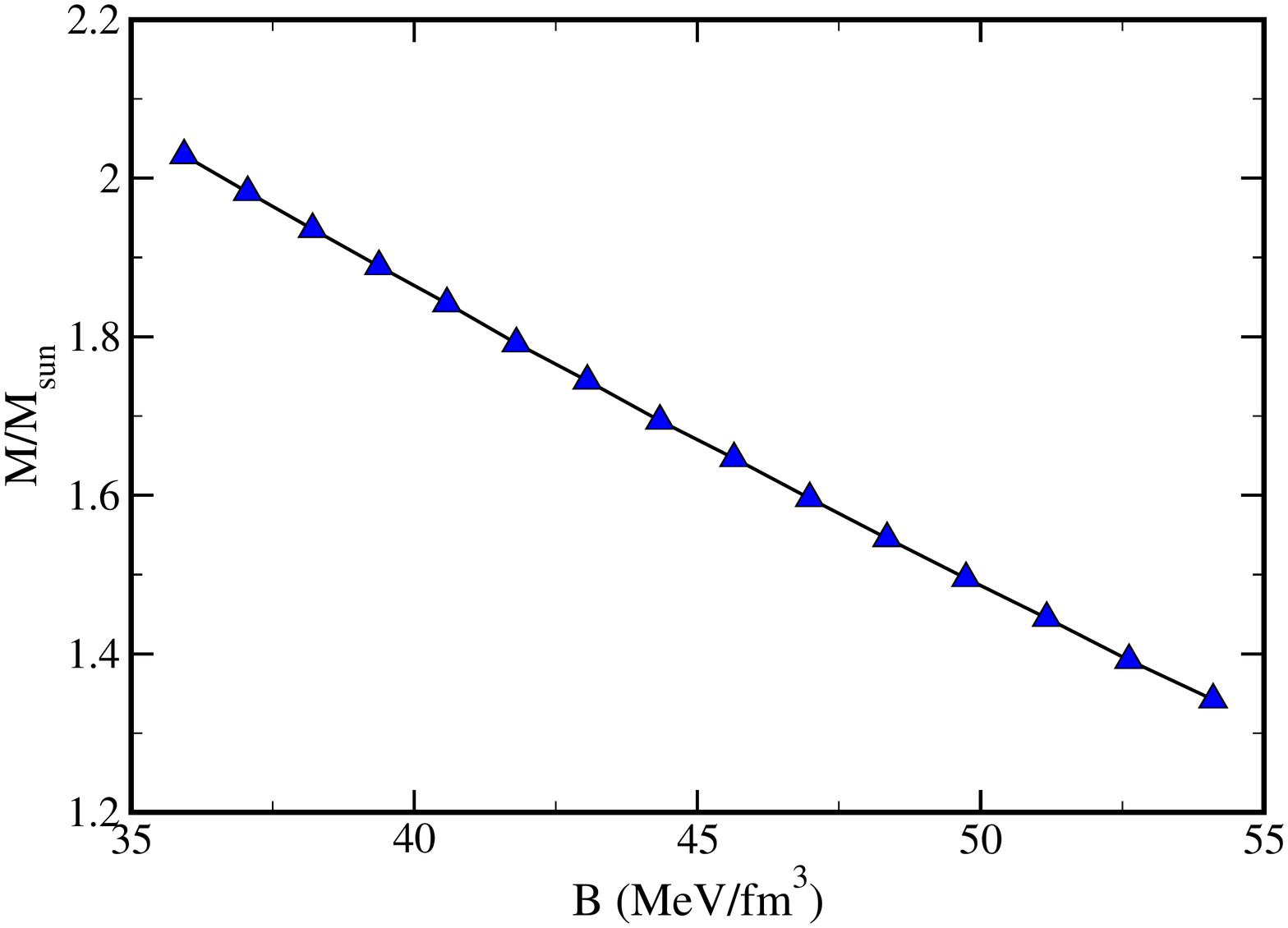}
\caption{\label{leftmode}Stellar mass vs Bag constant $B$ from Eq.\ref{eqn1} for a fixed central density of 3 x nuclear saturation density. $\mu_{\rm crit}$ is varied between 280 MeV to 320 MeV consistently with $B$, although values higher than $\mu_{\rm crit}$=310 MeV are excluded due to the requirement of self-bound quark matter.}
\end{minipage}\hspace{2pc}%
\begin{minipage}{18.5pc}
\includegraphics[width=19.8pc,height=14.2pc]{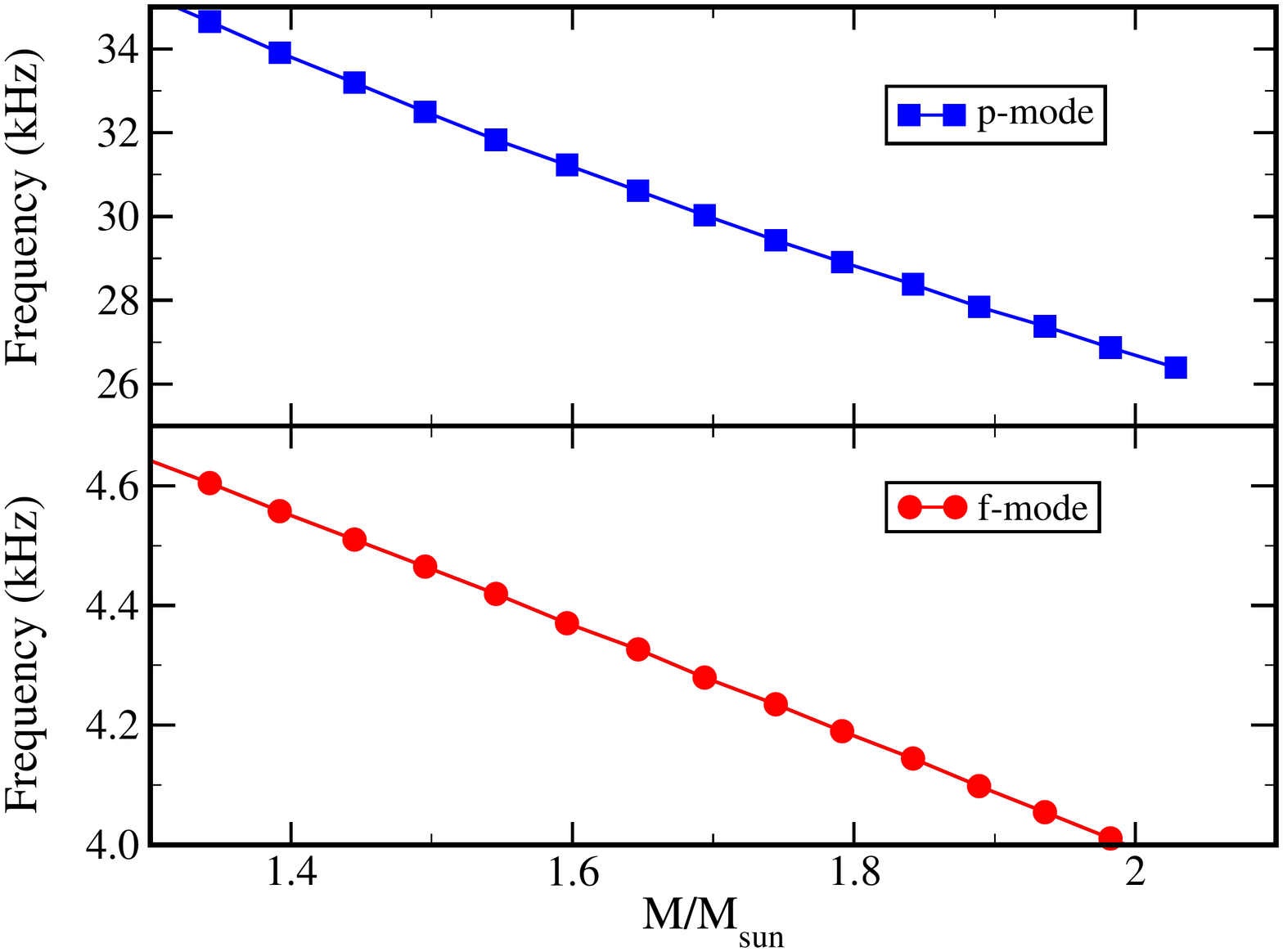}
\caption{\label{rightmode}The $f$ mode and first $p$ mode frequencies for homogeneous strange quark stars with $m_s$=100 MeV and $\mu_{\rm crit}$ varied self-consistently with $B$. Each pair of $f,p$ modes corresponding to a fixed mass represents a different EOS within the same family parametrized by Eq.~\ref{eqn1}. }
\end{minipage} 
\end{figure}

Next we compute the $f$ and $p$ modes for a homogeneous quark star having an
EOS defined as in Eq.~\ref{eqn1}, and study their dependence on the stellar mass. The mass of the strange quark, the
value of the critical quark chemical potential, and the interaction parameters $a_4$ are tunable features in the equation of state. Although the oscillation spectra are normally shown as a function of stellar mass for fixed $B$ (by changing central density), one should note that the Bag ``constant" is really an effective parameter of the model, and not an actual physical quantity.  We
therefore keep the value of quark mass fixed at 100 MeV and hold the central density constant, but vary the value of the Bag constant and $\mu_{\rm crit}$ self-consistently from the equation of state to generate different mass configurations. In Fig.\ref{leftmode}, we show how the stellar mass varies with the Bag constant. As seen in Fig.\ref{rightmode}, the trend of the $f$ modes is that as the bag constant is decreased (i.e, stellar mass increased), the frequencies gradually decrease, suggesting that the $f$ modes scale with the Bag constant rather than stellar mass. The $f$ mode lies in the 4-5 kHz range while the p1-mode lies in the 25-35 kHz range for $M\sim (1.2-2)M_{\odot}$. This range of frequencies for the $f$ and $p$ mode of strange quark stars is higher than found in the work of \cite{Flores:2013yqa} who used general relativistic fluid equations, so the differences could be attributed to our Newtonian approximations to the fluid equations, since their results for hadronic stars also show a similar difference from Newtonian models. In addition, each pair of data corresponding to a fixed mass in Fig.\ref{rightmode} represents a different EOS within the same family parametrized by Eq.~\ref{eqn1}, whereas the results of \cite{Flores:2013yqa} were plotted against mass keeping the EOS the same. The Newtonian approximation to the oscillation equations appears to yield mode frequencies for strange stars similar to the $n$=$2$ polytropic neutron star (see Table~\ref{polytrope-freq}), reflecting the fact that strange stars are more compact in comparison to the average neutron star. Furthermore, the longitudinal sound speed in quark stars is higher. These properties of the quark star are reflected in the slightly higher f and p-mode frequencies compared to, say, the n=1 polytrope for neutron stars.

\begin{figure}[h]
\begin{minipage}{18.8pc}
\includegraphics[width=19.8pc,height=16pc]{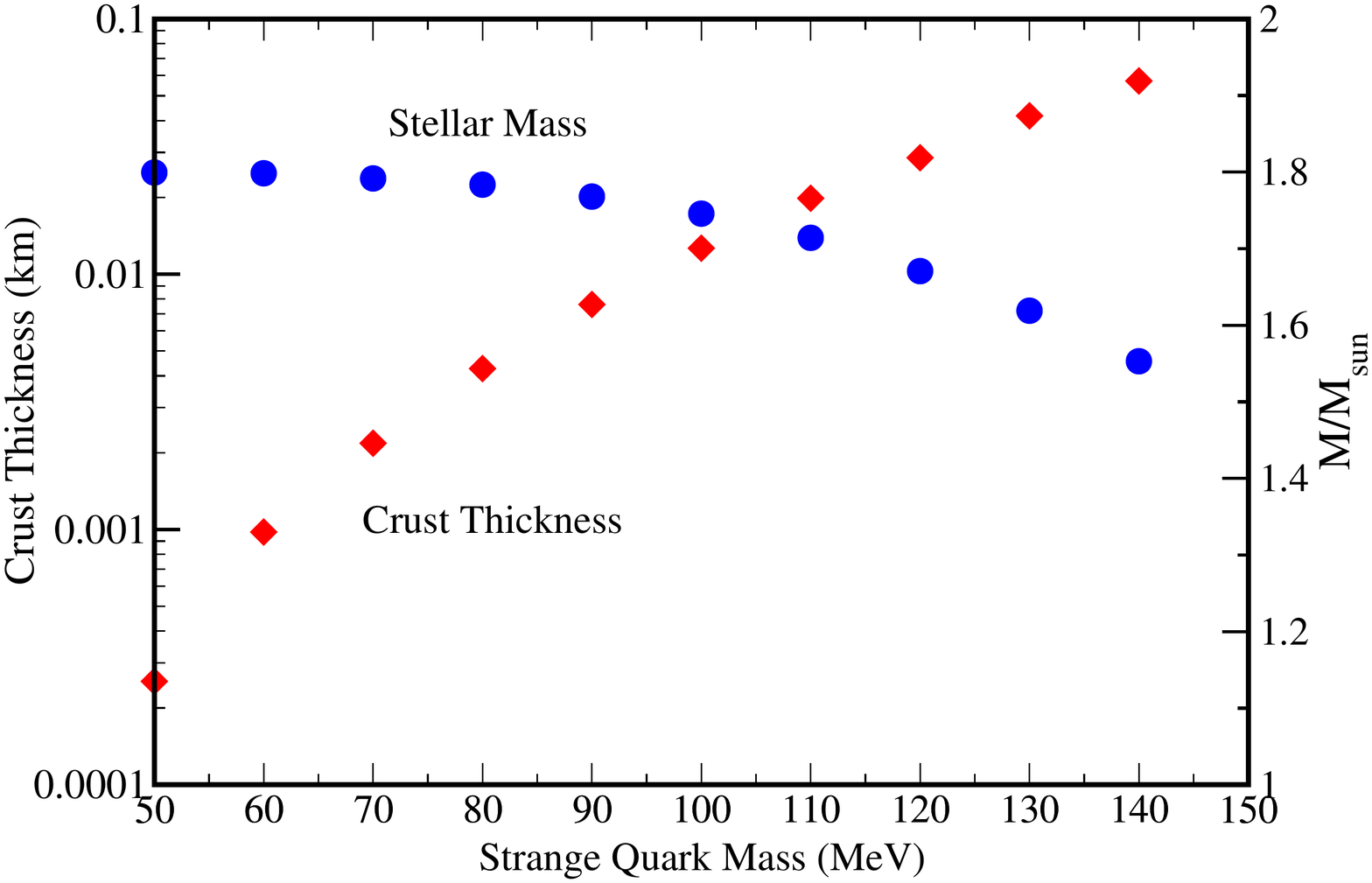}
\caption{\label{leftcrust}Crust thickness and stellar mass as a function of strange quark mass. The strangelet crust is constructed as a globally charge neutral phase atop the homogeneous phase which has zero pressure at $\mu_{\rm crit}$=310 MeV. }
\end{minipage}\hspace{2pc}%
\begin{minipage}{18.5pc}
\includegraphics[width=19.8pc,height=15.4pc]{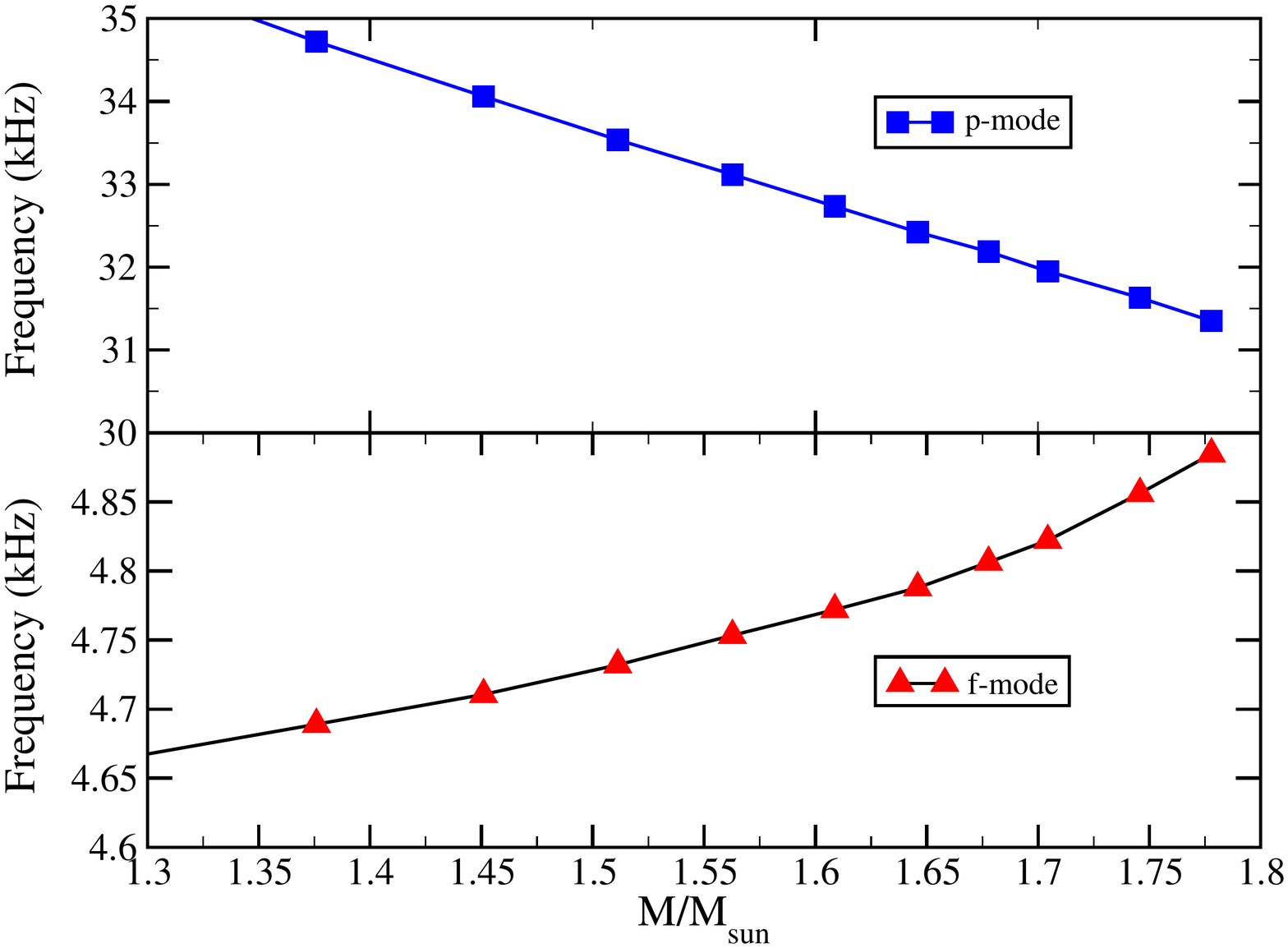}
\caption{\label{rightcrust}The $f$ mode and first $p$ mode frequencies for 2-component strange quark stars (core+strangelet crust) with $m_s$=100 MeV and $\mu_{\rm crit}$=310 MeV. Bag constant is held fixed as the central density is varied.}
\end{minipage} 
\end{figure}

The oscillation spectra for the two-component quark star (core+crust) was computed taking into account the shear modulus of the strangelet crust, which is estimated in~\cite{Watts:2006hk}. 
The effect of the crust on the $f$ modes is only a few Hz, since the mean density is hardly affected by adding the crust on top of the bare strange star. The $p$ modes are modified to a slightly larger extent than the $f$ mode, due to the shear modulus of the crust, but since the shear speed is numerically small compared to the longitudinal sound speed, the effect is not large (about 2\% change). An important EOS parameter for the strangelet crust's properties is the strange quark mass. In Fig.~\ref{leftcrust}, we show how the crust thickness depends on the strange quark mass for a fixed $\mu_{\rm crit}$, along with the corresponding stellar mass. In Fig.~\ref{rightcrust}, we show the $f$ and $p$-mode of the 2-component strange star with varying stellar mass keeping $\mu_{\rm crit}$ (and hence also the Bag constant) fixed, but changing the central density. The mode frequencies now have a similar trend to~\cite{Flores:2013yqa}, although the values differ due to reasons mentioned previously.

\section{Conclusions} 
In this proceedings article, we have presented some preliminary work on the spheroidal non-radial oscillation modes of strange quark stars that are composed of a homogeneous core and a crust of strangelets in a globally neutral phase (Gibbs construction). In the Newtonian approximation to the fluid equations, and assuming the fluid perturbations obey the same equation of state as the background fluid, the $f$ modes are in the range of 4-5 kHz while the p-modes are in the range of 25-35 kHz. Due to the small size of the crust and its low shear modulus, the impact on the modes from the case of a bare strange star is minimal. However, an extreme choice of parameters in the quark model can lead to a much thicker crust (hundreds of meters), and this affects the mode frequencies at almost the 10\% level. Including a distinct EOS for the perturbations, considering a crystalline superconducting crust, rotation effects, or high temperature can all change the results presented here. In this work, we have not presented the results for the shear modes ($s$ modes) or the interfacial modes ($i$ modes) that arise as a result of the mixed phase crust, and which would have much smaller frequencies than the $f,p$ modes. To resolve these modes, one needs a qualitatively new set of boundary conditions at the crust-core interface, since the energy density, and consequently the EOS varies sharply in that region. These modes are typically expected to lie between the $f,p$ modes and the core $g$ modes (few Hz to tens of Hz). Those findings and their comparison to the case of a neutron star with a crust or strange star with a nuclear crust, will be reported separately. The Newtonian approximation also appears to overestimate the mode frequencies when compared to including the perturbations in the gravitational potential~\cite{Sotani:2003zc}. A full general relativistic treatment along the lines in~\cite{Sotani:2003zc,Benhar:2004xg} is currently underway, which will yield the damping time and the amplitude of the modes for generating templates of the gravitational wave signal from pulsating strange stars. 

\section{Acknowledgments}
\vskip 0.1cm
P. J. thanks the organizers of the CSQCD V conference (Compact Stars in the QCD Phase Diagram V) held at the Gran Sasso Science Institute in L'Aquila, Italy, for their kind hospitality, and would also like to acknowledge fruitful discussions with G. Rupak (Mississippi State University, USA) and C. V\'asquez-Flores (Universidade Federal do ABC, Brazil). \\
\vskip 0.05cm
\noindent P. J. is supported by the U.S. National Science Foundation under Grant No. PHY 1608959.  \\J. A. acknowledges financial support from the Office of Research \& Sponsored Programs
at the California State University Long Beach.

\section{References}
\bibliography{jaikumar-refs}

\providecommand{\newblock}{}
\begin{thebibliography}{10}
\expandafter\ifx\csname url\endcsname\relax
  \def\url#1{{\tt #1}}\fi
\expandafter\ifx\csname urlprefix\endcsname\relax\def\urlprefix{URL }\fi
\providecommand{\eprint}[2][]{\url{#2}}

\bibitem{Abbott:2016blz}
Abbott B~P {\em et~al.\/} (Virgo, LIGO Scientific) 2016 {\em Phys. Rev.
  Lett.\/} {\bf 116} 061102 (\textit{Preprint} \eprint{1602.03837})

\bibitem{Abbott:2016nmj}
Abbott B~P {\em et~al.\/} (Virgo, LIGO Scientific) 2016 {\em Phys. Rev.
  Lett.\/} {\bf 116} 241103 (\textit{Preprint} \eprint{1606.04855})

\bibitem{Creighton:2003nm}
Creighton T 2003 {\em Class. Quant. Grav.\/} {\bf 20} S853--S869

\bibitem{Bauswein:2012ya}
Bauswein A, Janka H~T, Hebeler K and Schwenk A 2012 {\em Phys. Rev.\/} {\bf
  D86} 063001 (\textit{Preprint} \eprint{1204.1888})

\bibitem{Kokkotas:1999bd}
Kokkotas K~D and Schmidt B~G 1999 {\em Living Rev. Rel.\/} {\bf 2} 2
  (\textit{Preprint} \eprint{gr-qc/9909058})

\bibitem{Kokkotas:1999mn}
Kokkotas K~D, Apostolatos T~A and Andersson N 2001 {\em Mon. Not. Roy. Astron.
  Soc.\/} {\bf 320} 307--315 (\textit{Preprint} \eprint{gr-qc/9901072})

\bibitem{PenaGaray:2008qe}
Pena-Garay C and Serenelli A 2008  (\textit{Preprint} \eprint{0811.2424})

\bibitem{Sotani:2012qc}
Sotani H, Nakazato K, Iida K and Oyamatsu K 2012 {\em Phys. Rev. Lett.\/} {\bf
  108} 201101 (\textit{Preprint} \eprint{1202.6242})

\bibitem{Flores:2013yqa}
Flores C~V and Lugones G 2014 {\em Class. Quant. Grav.\/} {\bf 31} 155002
  (\textit{Preprint} \eprint{1310.0554})

\bibitem{Kojima:2002iv}
Kojima Y and Sakata K 2002 {\em Prog. Theor. Phys.\/} {\bf 108} 801--808
  (\textit{Preprint} \eprint{astro-ph/0209320})

\bibitem{Moraes:2014dra}
Moraes P~H~R~S and Miranda O~D 2014 {\em Mon. Not. Roy. Astron. Soc.\/} {\bf
  445} L11--L15 (\textit{Preprint} \eprint{1408.0929})

\bibitem{Chatziioannou:2015uea}
Chatziioannou K, Yagi K, Klein A, Cornish N and Yunes N 2015 {\em Phys. Rev.\/}
  {\bf D92} 104008 (\textit{Preprint} \eprint{1508.02062})

\bibitem{Fu:2017mcw}
Fu W, Bai Z and Liu Y 2017  (\textit{Preprint} \eprint{1701.00418})

\bibitem{Jaikumar:2005ne}
Jaikumar P, Reddy S and Steiner A~W 2006 {\em Phys. Rev. Lett.\/} {\bf 96}
  041101 (\textit{Preprint} \eprint{nucl-th/0507055})

\bibitem{Alford:2006bx}
Alford M~G, Rajagopal K, Reddy S and Steiner A~W 2006 {\em Phys. Rev.\/} {\bf
  D73} 114016 (\textit{Preprint} \eprint{hep-ph/0604134})

\bibitem{Sotani:2012ej}
Sotani H, Maruyama T and Tatsumi T 2013 {\em Nucl. Phys.\/} {\bf A906} 37--49
  (\textit{Preprint} \eprint{1207.4055})

\bibitem{1984PhRvD}
{Witten} E 1984 {\em Phys. Rev. D\/} {\bf 30} 272--285

\bibitem{Alford:2004pf}
Alford M, Braby M, Paris M~W and Reddy S 2005 {\em Astrophys. J.\/} {\bf 629}
  969--978 (\textit{Preprint} \eprint{nucl-th/0411016})

\bibitem{Fraga:2001id}
Fraga E~S, Pisarski R~D and Schaffner-Bielich J 2001 {\em Phys. Rev.\/} {\bf
  D63} 121702 (\textit{Preprint} \eprint{hep-ph/0101143})

\bibitem{Rupak:2012wk}
Rupak G and Jaikumar P 2013 {\em Phys. Rev.\/} {\bf C88} 065801
  (\textit{Preprint} \eprint{1209.4343})

\bibitem{Kastaun:2008jr}
Kastaun W 2008 {\em Phys. Rev.\/} {\bf D77} 124019 (\textit{Preprint}
  \eprint{0804.1151})

\bibitem{1988ApJ725M}
{McDermott} P~N, {van Horn} H~M and {Hansen} C~J 1988 {\em Astrophys. J.\/}
  {\bf 325} 725--748

\bibitem{Yoshida:2002vd}
Yoshida S and Lee U 2002 {\em Astron. Astrophys.\/} {\bf 395} 201--208
  (\textit{Preprint} \eprint{astro-ph/0210591})

\bibitem{Cox}
Cox J~P 1980  {\bf ed The Princeton University Press, New Jersey}

\bibitem{Andersson:1997rn}
Andersson N and Kokkotas K~D 1998 {\em Mon. Not. Roy. Astron. Soc.\/} {\bf 299}
  1059--1068 (\textit{Preprint} \eprint{gr-qc/9711088})

\bibitem{Watts:2006hk}
Watts A~L and Reddy S 2007 {\em Mon. Not. Roy. Astron. Soc.\/} {\bf 379} L63
  (\textit{Preprint} \eprint{astro-ph/0609364})

\bibitem{Sotani:2003zc}
Sotani H and Harada T 2003 {\em Phys. Rev.\/} {\bf D68} 024019
  (\textit{Preprint} \eprint{gr-qc/0307035})

\bibitem{Benhar:2004xg}
Benhar O, Ferrari V and Gualtieri L 2004 {\em Phys. Rev.\/} {\bf D70} 124015
  (\textit{Preprint} \eprint{astro-ph/0407529})

\end{thebibliography}

\end{document}